\documentclass[twocolumn,showpacs,aps,prl,superscriptaddress,floatfix,letterpaper]{revtex4}

\usepackage{graphicx}
\usepackage{dcolumn}
\usepackage{amsmath}
\usepackage{epsfig}
\usepackage{setspace}

\newcommand{\ifb}{\ensuremath{\mathrm{fb^{-1}}}}
\newcommand{\TeV}{\ensuremath{\mathrm{Te\kern -0.1em V}}}
\newcommand{\TeVc}{\ensuremath{\mathrm{Te\kern -0.1em V\!/}c}}
\newcommand{\TeVcc}{\ensuremath{\mathrm{Te\kern -0.1em V\!/}c^2}}
\newcommand{\GeV}{\ensuremath{\mathrm{Ge\kern -0.1em V}}}
\newcommand{\GeVc}{\ensuremath{\mathrm{Ge\kern -0.1em V\!/}c}}
\newcommand{\GeVcc}{\ensuremath{\mathrm{Ge\kern -0.1em V\!/}c^2}}
\newcommand{\MeV}{\ensuremath{\mathrm{Me\kern -0.1em V}}}
\newcommand{\MeVc}{\ensuremath{\mathrm{Me\kern -0.1em V\!/}c}}
\newcommand{\MeVcc}{\ensuremath{\mathrm{Me\kern -0.1em V\!/}c^2}}

\newcommand{\um}{\ensuremath{\mathrm{\mu m}}}

\newcommand{\babar}{\mbox{\slshape B\kern-0.1em{\small A}\kern-0.1em B\kern-0.1em{\small A\kern-0.2em R}}}

\def\mass{4143.0\pm2.9(\mathrm{stat})\pm1.2(\mathrm{syst})~\MeVcc }
\def\width{11.7^{+8.3}_{-5.0}(\mathrm{stat})\pm3.7(\mathrm{syst})~\MeVcc }
\def\yield{14\pm5 }
 
\def\massdifffit{1046.3\pm2.9~\MeVcc }
\def\widthfit{11.7^{+8.3}_{-5.0}~\MeVcc }
\def\massfit{4143.0\pm2.9~\MeVcc }

\begin{document}

\begin{flushleft}
\ \hfill\ \\
\ \hfill\ \\
\ \hfill\ \\
%\hfill Draft 2.1 \\
%\hfill xxxx-xxxx-xxxx 
\end{flushleft}

\title
{
{\large \bf 
Evidence for a Narrow  Near-Threshold Structure in the $J/\psi\phi$ Mass Spectrum  \\
in  $B^+\rightarrow  J/\psi\phi K^+$ Decays 
} 	
}

\affiliation{Institute of Physics, Academia Sinica, Taipei, Taiwan 11529, Republic of China} 
\affiliation{Argonne National Laboratory, Argonne, Illinois 60439} 
\affiliation{University of Athens, 157 71 Athens, Greece} 
\affiliation{Institut de Fisica d'Altes Energies, Universitat Autonoma de Barcelona, E-08193, Bellaterra (Barcelona), Spain} 
\affiliation{Baylor University, Waco, Texas  76798} 
\affiliation{Istituto Nazionale di Fisica Nucleare Bologna, $^y$University of Bologna, I-40127 Bologna, Italy} 
\affiliation{Brandeis University, Waltham, Massachusetts 02254} 
\affiliation{University of California, Davis, Davis, California  95616} 
\affiliation{University of California, Los Angeles, Los Angeles, California  90024} 
\affiliation{University of California, San Diego, La Jolla, California  92093} 
\affiliation{University of California, Santa Barbara, Santa Barbara, California 93106} 
\affiliation{Instituto de Fisica de Cantabria, CSIC-University of Cantabria, 39005 Santander, Spain} 
\affiliation{Carnegie Mellon University, Pittsburgh, PA  15213} 
\affiliation{Enrico Fermi Institute, University of Chicago, Chicago, Illinois 60637}
\affiliation{Comenius University, 842 48 Bratislava, Slovakia; Institute of Experimental Physics, 040 01 Kosice, Slovakia} 
\affiliation{Joint Institute for Nuclear Research, RU-141980 Dubna, Russia} 
\affiliation{Duke University, Durham, North Carolina  27708} 
\affiliation{Fermi National Accelerator Laboratory, Batavia, Illinois 60510} 
\affiliation{University of Florida, Gainesville, Florida  32611} 
\affiliation{Laboratori Nazionali di Frascati, Istituto Nazionale di Fisica Nucleare, I-00044 Frascati, Italy} 
\affiliation{University of Geneva, CH-1211 Geneva 4, Switzerland} 
\affiliation{Glasgow University, Glasgow G12 8QQ, United Kingdom} 
\affiliation{Harvard University, Cambridge, Massachusetts 02138} 
\affiliation{Division of High Energy Physics, Department of Physics, University of Helsinki and Helsinki Institute of Physics, FIN-00014, Helsinki, Finland} 
\affiliation{University of Illinois, Urbana, Illinois 61801} 
\affiliation{The Johns Hopkins University, Baltimore, Maryland 21218} 
\affiliation{Institut f\"{u}r Experimentelle Kernphysik, Universit\"{a}t Karlsruhe, 76128 Karlsruhe, Germany} 
\affiliation{Center for High Energy Physics: Kyungpook National University, Daegu 702-701, Korea; Seoul National University, Seoul 151-742, Korea; Sungkyunkwan University, Suwon 440-746, Korea; Korea Institute of Science and Technology Information, Daejeon, 305-806, Korea; Chonnam National University, Gwangju, 500-757, Korea} 
\affiliation{Ernest Orlando Lawrence Berkeley National Laboratory, Berkeley, California 94720} 
\affiliation{University of Liverpool, Liverpool L69 7ZE, United Kingdom} 
\affiliation{University College London, London WC1E 6BT, United Kingdom} 
\affiliation{Centro de Investigaciones Energeticas Medioambientales y Tecnologicas, E-28040 Madrid, Spain} 
\affiliation{Massachusetts Institute of Technology, Cambridge, Massachusetts  02139} 
\affiliation{Institute of Particle Physics: McGill University, Montr\'{e}al, Qu\'{e}bec, Canada H3A~2T8; Simon Fraser University, Burnaby, British Columbia, Canada V5A~1S6; University of Toronto, Toronto, Ontario, Canada M5S~1A7; and TRIUMF, Vancouver, British Columbia, Canada V6T~2A3} 
\affiliation{University of Michigan, Ann Arbor, Michigan 48109} 
\affiliation{Michigan State University, East Lansing, Michigan  48824}
\affiliation{Institution for Theoretical and Experimental Physics, ITEP, Moscow 117259, Russia} 
\affiliation{University of New Mexico, Albuquerque, New Mexico 87131} 
\affiliation{Northwestern University, Evanston, Illinois  60208} 
\affiliation{The Ohio State University, Columbus, Ohio  43210} 
\affiliation{Okayama University, Okayama 700-8530, Japan} 
\affiliation{Osaka City University, Osaka 588, Japan} 
\affiliation{University of Oxford, Oxford OX1 3RH, United Kingdom} 
\affiliation{Istituto Nazionale di Fisica Nucleare, Sezione di Padova-Trento, $^z$University of Padova, I-35131 Padova, Italy} 
\affiliation{LPNHE, Universite Pierre et Marie Curie/IN2P3-CNRS, UMR7585, Paris, F-75252 France} 
\affiliation{University of Pennsylvania, Philadelphia, Pennsylvania 19104}
\affiliation{Istituto Nazionale di Fisica Nucleare Pisa, $^{aa}$University of Pisa, $^{bb}$University of Siena and $^{cc}$Scuola Normale Superiore, I-56127 Pisa, Italy} 
\affiliation{University of Pittsburgh, Pittsburgh, Pennsylvania 15260} 
\affiliation{Purdue University, West Lafayette, Indiana 47907} 
\affiliation{University of Rochester, Rochester, New York 14627} 
\affiliation{The Rockefeller University, New York, New York 10021} 
\affiliation{Istituto Nazionale di Fisica Nucleare, Sezione di Roma 1, $^{dd}$Sapienza Universit\`{a} di Roma, I-00185 Roma, Italy} 

\affiliation{Rutgers University, Piscataway, New Jersey 08855} 
\affiliation{Texas A\&M University, College Station, Texas 77843} 
\affiliation{Istituto Nazionale di Fisica Nucleare Trieste/Udine, I-34100 Trieste, $^{ee}$University of Trieste/Udine, I-33100 Udine, Italy} 
\affiliation{University of Tsukuba, Tsukuba, Ibaraki 305, Japan} 
\affiliation{Tufts University, Medford, Massachusetts 02155} 
\affiliation{Waseda University, Tokyo 169, Japan} 
\affiliation{Wayne State University, Detroit, Michigan  48201} 
\affiliation{University of Wisconsin, Madison, Wisconsin 53706} 
\affiliation{Yale University, New Haven, Connecticut 06520} 
\author{T.~Aaltonen}
\affiliation{Division of High Energy Physics, Department of Physics, University of Helsinki and Helsinki Institute of Physics, FIN-00014, Helsinki, Finland}
\author{J.~Adelman}
\affiliation{Enrico Fermi Institute, University of Chicago, Chicago, Illinois 60637}
\author{T.~Akimoto}
\affiliation{University of Tsukuba, Tsukuba, Ibaraki 305, Japan}
\author{B.~\'{A}lvarez~Gonz\'{a}lez$^t$}
\affiliation{Instituto de Fisica de Cantabria, CSIC-University of Cantabria, 39005 Santander, Spain}
\author{S.~Amerio$^z$}
\affiliation{Istituto Nazionale di Fisica Nucleare, Sezione di Padova-Trento, $^z$University of Padova, I-35131 Padova, Italy} 

\author{D.~Amidei}
\affiliation{University of Michigan, Ann Arbor, Michigan 48109}
\author{A.~Anastassov}
\affiliation{Northwestern University, Evanston, Illinois  60208}
\author{A.~Annovi}
\affiliation{Laboratori Nazionali di Frascati, Istituto Nazionale di Fisica Nucleare, I-00044 Frascati, Italy}
\author{J.~Antos}
\affiliation{Comenius University, 842 48 Bratislava, Slovakia; Institute of Experimental Physics, 040 01 Kosice, Slovakia}
\author{G.~Apollinari}
\affiliation{Fermi National Accelerator Laboratory, Batavia, Illinois 60510}
% ...

\collaboration{CDF Collaboration\footnote{With visitors from $^a$University of Massachusetts Amherst, Amherst, Massachusetts 01003,
$^b$Universiteit Antwerpen, B-2610 Antwerp, Belgium, 
$^c$University of Bristol, Bristol BS8 1TL, United Kingdom,
$^d$Chinese Academy of Sciences, Beijing 100864, China, 
$^e$Istituto Nazionale di Fisica Nucleare, Sezione di Cagliari, 09042 Monserrato (Cagliari), Italy,
$^f$University of California Irvine, Irvine, CA  92697, 
$^g$University of California Santa Cruz, Santa Cruz, CA  95064, 
$^h$Cornell University, Ithaca, NY  14853, 
$^i$University of Cyprus, Nicosia CY-1678, Cyprus, 
$^j$University College Dublin, Dublin 4, Ireland,
$^k$University of Edinburgh, Edinburgh EH9 3JZ, United Kingdom, 
$^l$University of Fukui, Fukui City, Fukui Prefecture, Japan 910-0017
$^m$Kinki University, Higashi-Osaka City, Japan 577-8502
$^n$Universidad Iberoamericana, Mexico D.F., Mexico,
$^o$University of Iowa, Iowa City, IA  52242,
$^p$Queen Mary, University of London, London, E1 4NS, England,
$^q$University of Manchester, Manchester M13 9PL, England, 
$^r$Nagasaki Institute of Applied Science, Nagasaki, Japan, 
$^s$University of Notre Dame, Notre Dame, IN 46556,
$^t$University de Oviedo, E-33007 Oviedo, Spain, 
$^u$Texas Tech University, Lubbock, TX  79609, 
$^v$IFIC(CSIC-Universitat de Valencia), 46071 Valencia, Spain,
$^w$University of Virginia, Charlottesville, VA  22904,
$^x$Bergische Universit\"at Wuppertal, 42097 Wuppertal, Germany,
$^{ff}$On leave from J.~Stefan Institute, Ljubljana, Slovenia, 
}}
\noaffiliation

\date{\today}% It is always \today, today, but you may specify any date with \date.

\begin{abstract}

Evidence  is reported for a narrow  structure near the  $J/\psi\phi$ threshold   
in exclusive $B^+\rightarrow J/\psi\phi K^+$ decays produced in $\bar{p} p $
collisions at $\sqrt{s}=1.96~\TeV$. 
A signal of $\yield$ events, with statistical significance in excess of  
3.8 standard deviations, is observed in a data sample corresponding to an 
integrated luminosity  of  $2.7~\ifb$,  collected by the CDF~II detector. 
The mass and natural width of the structure are measured to be $\mass$ and $\width$.

\end{abstract}

\pacs{
14.40.Gx, %  Properties of specific particles:  Mesons with S=C=B=0, mass>2.5 GeV (including quarkonia)
13.25.Gv,  %  Hadronic decays of mesons: Decays of J/psi, Upsilon, and other quarkonia 
12.39.Mk %  Glueball and nonstandard multi-quark/gluon states 
 }     % PACS, the Physics and Astronomy

%\pacs{12.20.-m, 12.38.Aw, 12.40.Vv}
% PACS, the Physics and Astronomy Classification Scheme.

\maketitle

%\begin{linenumbers}

Heavy quarkonium spectroscopy provides insight into strong interactions that  
are not precisely predictable by QCD theory. The recently discovered states   
that have charmonium-like decay modes~\cite{x3872discovery,y3940,y4260discovery,y4320} 
but are difficult to place in the overall charmonium system
have introduced new challenges.  The possible interpretations beyond quark-antiquark 
states ($q\bar{q}$)  
such as hybrid ($q\bar{q}g$) and four-quark states ($q\bar{q}q\bar{q}$)
have revitalized  interest in exotic mesons in the charm 
sector~\cite{conventional,hybrid,molecule,fourquark}.   
An important tool in unraveling the nature of the states 
in the charmonium-mass region is the  exploration 
of states in diverse channels.
First, the $J/\psi\phi$ final state, with positive C-parity, 
and two $\mathrm{J^{PC}}=1^{--}$ vector mesons (VV),   
is a good channel for an exotic meson search. 
The discovery of the $X(3872)$ (proposed as   
a four-quark state ($c\bar{c}q\bar{q}$)~\cite{molecule,fourquark})
 and $Y(3930)$~\cite{y3940}, both decaying into  
VV~\cite{x3872jpc}, suggests searching for
 other possible VV states~\cite{ric}.    
Second, the observation of $Y(3930)$ near the $J/\psi\omega$ threshold 
motivates searches for similar phenomena near the $J/\psi\phi$ threshold. 
Third, the observation of the $Y(4260)$, a potential hybrid candidate~\cite{hybrid},
leads to an expectation of a triplet of hybrid states
$J^{PC} =(0, 1, 2)^{-+}$ to lie nearby 
in mass~\cite{estiacharm07,hybridfromb}, which would be   
accessible in the $J/\psi\phi$ channel.
Finally, other possibilities such as glueballs~\cite{hybridfromb} and  
nuclear-bound quarkonium~\cite{stanquarkonium} 
also motivate  a search in this channel. 
The $J/\psi \phi$ channel is accessible in the decay mode
$B^+\to J/\psi \phi K^+$, which  has been observed~\cite{cleojpsiphi}. 
However,  to date no
results have been reported for substructure in the $J/\psi \phi$ channel.

In this Letter, we report an investigation of the $J/\psi\phi$ system produced in exclusive 
$B^+\rightarrow J/\psi\phi K^+$ decays with $J/\psi \rightarrow \mu^+ \mu^-$ 
and $\phi\rightarrow K^+K^-$. The search in exclusive $B^+$ decays is  
more sensitive than an inclusive search since 
the additional $B^+$ mass constraint on the $J/\psi\phi K^+$ system helps to reduce background.
This analysis is based on a data sample  of $\bar{p} p $
collisions at $\sqrt{s}=1.96~\TeV$ with an integrated luminosity of 2.7 $\ifb$
collected by the CDF II detector at the Tevatron. 
Charge conjugate modes are included implicitly in this Letter.

The CDF II detector has been described in detail elsewhere~\cite{cdfii}. 
The important components  for this analysis include the tracking,  
muon, and   time-of-flight (TOF) systems. 
The tracking system is composed of a  silicon-strip 
vertex detector (SVX) surrounded by an 
open-cell  drift chamber system called the central outer tracker  (COT)
located inside a solenoid with a 1.4 T magnetic field.
The COT and SVX are used for the measurement of charged-particle
  trajectories and vertex positions.  In addition, the COT provides ionization energy loss
  information, ${dE/dx}$, used for particle identification (PID), while the TOF system provides
  complementary PID information.
The muon system is located radially outside the electromagnetic and hadronic calorimeters
and consists of two sets of drift chambers and scintillation counters.
The central part of the muon system covers the pseudorapidity region 
$\left| \eta \right| \le  0.6$ and detects muons 
with $p_T~\ge~1.4~\GeVc$~\cite{eta}, and the second part 
covers the region  $0.6< \left| \eta \right| <  1.0$ and 
detects muons with $p_T~\ge~2.0~\GeVc$.

In this analysis,
$J/\psi\to\mu^+\mu^-$ events are recorded 
using a dedicated three-level dimuon trigger. The first trigger level 
requires two muon candidates with matching tracks in the COT and muon systems. 
The second level applies additional  kinematic requirements to the muon pair candidate.
The third level requires the invariant mass of the $\mu^+\mu^-$ pair to be 
within the range of $2.7$  to $4.0$ $\GeVcc$.

Offline reconstruction of $B^+ \rightarrow J/\psi \phi K^+$ candidates uses only
tracks that pass standard CDF quality requirements and which have been corrected for
ionization energy loss for the muon or kaon hypothesis, as appropriate.
The  $B^+\rightarrow J/\psi\phi K^+$ candidates are reconstructed 
by combining a $J/\psi\to\mu^+\mu^-$ candidate, a $\phi\rightarrow K^+K^-$ candidate,  
and an additional charged track.
All five tracks must form a good quality 3D vertex, using {\it a priori} 
requirements typical for $B$ hadron reconstruction at CDF~\cite{cdfbslifetime}.
Preliminary event selection  requires a $J/\psi$ candidate reconstructed
using opposite-sign muon candidates and a  $\phi$ candidate formed from
opposite-sign tracks to which we assign the kaon mass.  
Masses of vector meson candidates must lie within 
$\pm50$ \MeVcc~ of the  $J/\psi$ mass for muons or $\pm7$ \MeVcc~ of the $\phi$ mass for
kaons.
In the final $B^+$ reconstruction the $J/\psi$ is mass constrained, and the $B^+$ 
candidates must have $p_T > 4$ \GeVc.

\begin{figure}
\includegraphics[width=9.0cm,height=4.3cm]{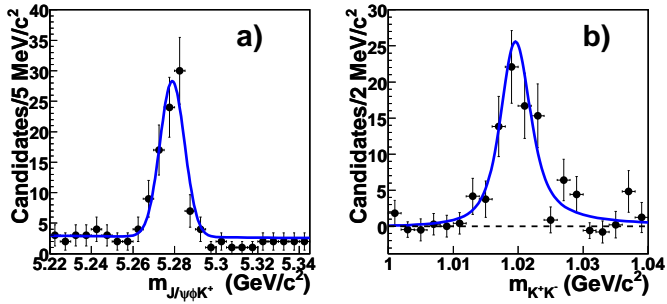}
\caption{\label{jpsiphik} 
(a) The mass distribution of $J/\psi\phi K^+$; 
the solid line is a fit to the data with a Gaussian signal 
function and flat background function.
(b) The $B^+$ sideband-subtracted mass distribution of  $K^+K^-$ 
without the $\phi$ mass window requirement.
The solid curve is a $P$-wave relativistic Breit-Wigner fit to the data.
}%end caption
\end{figure}

To suppress combinatorial background, we use ${dE/dx}$ and 
$\mathrm{TOF}$ information to identify all three kaons in the final state.
The information is summarized in a log-likelihood ratio ($LLR$), which 
reflects how well a candidate track can be positively identified 
as a kaon relative to other hadrons~\cite{bsmix}.
In addition, we require a minimum ${L_{xy}(B^+)}$ for the 
$B^+\rightarrow J/\psi\phi K^+$ candidate,
where ${L_{xy}(B^+)}$ is the projection onto ${\vec{p}_T(B^+)}$ 
of the vector connecting the primary vertex to the ${B^+}$ decay vertex.   
The primary vertex is determined for each event using  
prompt tracks.

The  ${L_{xy}(B^+)}$ and   ${LLR}$ requirements for $B^+\rightarrow J/\psi\phi K^+$ 
are then chosen to 
maximize $\mathit{S/\sqrt{S+B}}$,  
where $\mathit{S}$ is the number of $B^+\rightarrow J/\psi\phi K^+$ signal events 
and $\mathit{B}$ is the number of background events in the $J/\psi\phi K^+$ 
mass range of $5.0$ to $5.6$~GeV/$c^2$ in the data.  The values of $\mathit{S}$ 
and $\mathit{B}$ are determined 
from an unbinned log-likelihood fit to the mass spectrum of $J/\psi\phi K^+$,  
for a given  set of values of ${L_{xy}(B^+)}$ and  ${LLR}$.
A Gaussian function is used to represent the $B^+\rightarrow J/\psi\phi K^+$ signal, where 
the mean value of the Gaussian 
is fixed to the $B^+$  world-average mass value~\cite{PDG}. The $B^+$ mass resolution 
is fixed to  the value 5.9 \MeVcc~obtained from Monte Carlo (MC) simulation~\cite{cdfbc}. 
A linear function is used to model the background in the fit.
The requirements  obtained by  maximizing $\mathit{S/\sqrt{S+B}}$ are ${L_{xy}(B^+)}>500 ~\um$ 
and  ${LLR}>0.2$. 
In order to study the efficiency of the ${L_{xy}(B^+)}$ and ${LLR}$ selections, we also 
reconstruct $B^+\rightarrow J/\psi K^+$ and  $B^0_s\rightarrow J/\psi \phi$ as control 
channels. We  select approximately 
50 000  $B^+\rightarrow J/\psi K^+$ and 3000 $B^0_s\rightarrow J/\psi \phi$ events
by applying similar requirements as for the $J/\psi\phi K^+$ channel but 
without the ${L_{xy}(B^+)}$ and   ${LLR}$ requirements. 
The efficiency for PID with the ${LLR}>0.2$ requirement 
is approximately 80\% per kaon and  is reasonably flat as a function of 
kaon $p_T$; the efficiency for  ${L_{xy}(B^+)}>500 ~\um$ is approximately 60\%,  
based on the $B^+ \rightarrow J/\psi K^+$ control sample.

The invariant mass of $J/\psi\phi K^+$ after the 
${L_{xy}(B^+)}$ and ${LLR}$ requirements and $J/\psi$ and $\phi$ mass window requirements is shown in 
Fig.~\ref{jpsiphik}(a). 
A fit with a Gaussian signal function and a 
flat background function  to the mass spectrum of $J/\psi\phi K^+$  
returns a $B^+$ signal of $75\pm10(\mathrm{stat})$ events.
We select  $B^+$ signal candidates with a mass 
within  3$\sigma$ (17.7 \MeVcc) of the nominal $B^+$ mass; 
the purity of the $B^+$ signal in that mass window is approximately 80\%.

The combinatorial background under the $B^+$ peak includes
$B$ hadron decays such as $B^0_s \rightarrow \psi(2S)\phi \rightarrow J/\psi \pi^+\pi^-\phi$,
in which the pions are misidentified as kaons.  However,
background events with misidentified kaons cannot
yield a Gaussian peak at the $B^+$ mass consistent with the
5.9 \MeVcc mass resolution.  The kinematics are such
that for the hypothesis $B^+ \rightarrow J/\psi K^+K^-K^+$, only events with
real kaons can produce the observed Gaussian signal.
Thus, with the $B^+$ mass window selection the sample
consists of real $B^+ \rightarrow J/\psi K^+K^-K^+$ decays over a small
combinatorial background.

Figure~\ref{jpsiphik}(b) shows the invariant mass distribution of $K^+ K^-$ pairs 
from $\mu^+\mu^-K^+K^- K^+$ candidates within $\pm 3 \sigma$ of the nominal $B^+$ mass.
The spectrum shown in this figure has had the sidebands subtracted, but
the $\phi$ mass window selection has not been applied.
By fitting the $K^+K^-$ mass spectrum to a $P$-wave relativistic 
Breit-Wigner (BW) function~\cite{pbw} convoluted with 
a Gaussian resolution function with the rms fixed to 1.3 \MeVcc~obtained from 
simulation, we obtain a mass of
$1019.6\pm 0.3$ \MeVcc~ and a width 
of $3.84\pm   0.65$ \MeVcc with $\chi^2$ probability of 28\%,  
consistent with the world-average values for the $\phi$ meson~\cite{PDG}.
The good fit indicates that after the
$\pm 7$ $\MeVcc$ selection on the $\phi$ mass window, 
the $B^+ \rightarrow J/\psi K^+K^-K^+$ final state is well described 
as $J/\psi \phi K^+$, with negligible contributions from
$J/\psi f_0(980) K^+$ or $J/\psi K^+K^-K^+$ phase space.

\begin{figure}
\includegraphics[width=9.0cm,height=4.3cm]{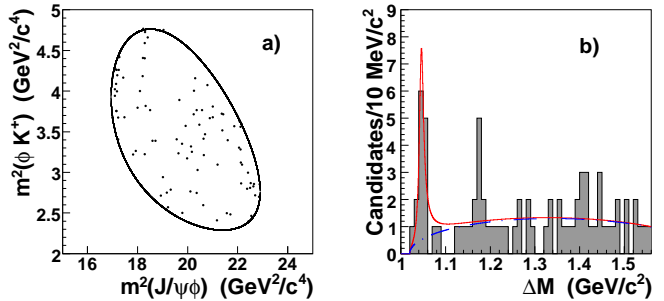}
\caption{\label{dalitz} 
(a) The Dalitz plot of ${m^2(\phi K^+)}$ 
versus ${m^2(J/\psi \phi)}$  in the  $B^+$ 
mass window. The boundary shows the kinematic allowed region. 
(b) The mass difference, $\Delta M$, between $\mu^+\mu^-K^+K^-$ and $\mu^+\mu^-$,  
 in the $B^+$ mass window. 
The dash-dotted curve is 
the background contribution and the red solid curve is the total unbinned fit. 
}%end caption
\end{figure}

We examine the effects of detector acceptance and selection
  requirements using 
$B^+ \rightarrow J/\psi \phi K^+$ MC events simulated  by phase space distributions. 
The MC events are smoothly distributed in the Dalitz plot and 
in the $J/\psi \phi$ mass spectrum. 
Figure~\ref{dalitz}(a) shows the Dalitz plot of ${m^2(\phi K^+)}$ 
versus ${m^2(J/\psi \phi)}$,   
and  Fig.~\ref{dalitz}(b) shows 
the mass difference, $\Delta M= m(\mu^+\mu^-K^+K^-)- m(\mu^+\mu^-)$, for events in 
the $B^+$ mass window in our data sample. 
We examine the enhancement in the $\Delta M$ spectrum just above $J/\psi\phi$ threshold,
using 73 events with  $\Delta M<1.56$ \GeVcc.
We exclude the high mass part of the spectrum to avoid 
combinatorial backgrounds  from misidentified 
$B^0_s\rightarrow \psi(2S) \phi\rightarrow (J/\psi\pi^+\pi^-)\phi$ decays.

We model the enhancement by an $S$-wave relativistic BW 
function~\cite{sbw} convoluted with 
a Gaussian resolution function with the rms fixed to 1.7 \MeVcc~obtained from MC,  
and use three--body phase space~\cite{PDG} to describe the background shape.
An unbinned likelihood fit to the $\Delta M$ distribution,   
as shown in Fig.~\ref{dalitz}(b), returns a yield of $\yield$ events,   
a $\Delta M$ of $\massdifffit$, and a width of $\widthfit$.
We also fit the $\Delta M$ distribution to
a single Gaussian with rms given by the mass 
resolution (1.7 \MeVcc), plus phase space background,   
to test the hypothesis that the structure 
has zero width. 
The statistical significance for a non-zero width determined by 
the log-likelihood ratio 
between these two fits
is 3.4$\sigma$, indicating a strong decay for this structure.

We use the log-likelihood ratio of $-2{\mathrm{ln}}(\mathcal{L}_0/\mathcal{L}_{{max}})$
to determine the significance of the structure at the $J/\psi \phi$ threshold, 
where $\mathcal{L}_0$ and $\mathcal{L}_{{max}}$ are the likelihood 
values for the null  hypothesis fit and signal hypothesis fit.
The $\sqrt{-2{\mathrm{ln}}(\mathcal{L}_0/\mathcal{L}_{{max}})}$ value 
is 5.3 for a pure three--body phase space 
background shape assumption. 
  To estimate the probability that background fluctuations alone
  would give rise to signals as significant as that seen in the data, 
  we simulate $\Delta M$ spectra based on the background distribution
  alone, and search for the most significant fluctuation in
  each spectrum in the mass range of 1.02 to 1.56 \GeVcc, with widths
  in the range of 1.7 (resolution) to 120 \MeVcc~(ten times of the observed width).  
From these spectra we obtain the  
  distribution for the quantity $-2{\mathrm{ln}}(\mathcal{L}_0/\mathcal{L}_{{max}})$ 
  in pure  background samples, and compare this with the signal in the data.
We performed a total of 3.1 million 
simulations and found  29 trials with a $-2{\mathrm{ln}}(\mathcal{L}_0/\mathcal{L}_{{max}})$ value 
greater than or equal to the value obtained in the data. 
The resulting $p$-value  
is $9.3\times 10^{-6}$, corresponding to a significance of 4.3$\sigma$.
Thus, the significance is decreased from a simple estimate of 
   $5.3\sigma$ to 4.3$\sigma$ by taking into account the absence of a prior
   prediction for the mass and 
width~\footnote{Monte Carlo studies indicate that fits with small numbers of signal
   events tend to return statistical uncertainties that underestimate
   the true uncertainties, an effect that asymptotically decreases
   as the sample size becomes larger. We have not corrected for such an effect.  Such an
underestimate of the fitted parameter uncertainties does not influence the
evaluation of the signal significance, which depends only on the
background fluctuation probability.}.

In the analysis described above, we assumed that
       the backgrounds  to the BW signal from both 
       $B^+ \rightarrow J/\psi \phi K^+$ decays  and combinatorial events 
       in the $B^+$ mass window are described
       by three-body phase space. 
After making the $\Delta M < 1.56$ \GeVcc~ selection, we perform a
fit to the $J/\psi \phi K^+$ mass spectrum with a Gaussian $B^+$ signal and
an empirical linear background shape; from this fit we determine that the
73 events within the final $B^+$ mass window include 
$15\pm1$ combinatorial background events by scaling it from the entire
$J/\psi\phi K^+$ mass range. 
We model the
       $B^+$ events using three-body phase space as above, but use 
       a flat spectrum to describe the combinatorial
       events. This increases the average background level
       at small $\Delta M$ and reduces  the yield by one event. 
The $\sqrt{-2{\mathrm{ln}}(\mathcal{L}_0/\mathcal{L}_{{max}})}$ value with this modeling of
       background 
is 4.8.
We performed a total of 1.1 million simulations and 
found  99 trials with a $-2{\mathrm{ln}}(\mathcal{L}_0/\mathcal{L}_{{max}})$ value 
greater than or equal to the value we obtained in the data. The $p$-value determined by this 
MC simulation is $9.0\times 10^{-5}$, about 3.8$\sigma$ significance.

The mass of this structure is $\massfit$ after including the world-average $J/\psi$ mass. 
To study the systematic uncertainties of the mass and width, we repeat the fit to  
the $\Delta M$ distribution while varying  
the background shapes as described above,  
and separately switching to a non--relativistic BW for signal. 
The largest deviation from the nominal values 
are 1.2 $\MeVcc$ for $\Delta M$ and 3.7 $\MeVcc$ for the width. 
Therefore we assign a systematic uncertainty of 
 1.2 $\MeVcc$ to the mass and 3.7 $\MeVcc$ to the width.

There is a small cluster of events approximately one pion mass higher than 
the first structure, located around 1.18 \GeVcc~in Fig.~\ref{dalitz}(b).
However, the statistical significance of this cluster is less than 3$\sigma$.
To investigate possible reflections, we examine the Dalitz plot and 
projections into  $\phi K^+$ and $J/\psi K^+$ spectrum.  
We find no evidence for any other structure in the $\phi K^+$ and $J/\psi K^+$ 
spectrum; the only structure [{\it i.e.} $K_2(1770)$] that has been claimed
in the  $\phi K^+$ spectrum by previous experiments is too broad 
to alter our analysis~\cite{k1770}.

In summary, the large sample of
$B^+ \rightarrow J/\psi \phi K^+$ decays (75 events)  enables us 
to search for structure in the $J/\psi\phi$ mass spectrum,  
and we find  evidence for a narrow structure near the $J/\psi\phi$
threshold with a significance in excess of 3.8$\sigma$.
Assuming an $S$-wave relativistic BW, the mass and width of this structure, 
including systematic  uncertainties,  are measured to be $\mass$ 
and $\width$, respectively.   
It is well above the threshold for open charm decays,   
so a $c\bar{c}$ charmonium meson with this mass would be expected to decay into
an open charm pair dominantly and to have a tiny branching fraction into 
$J/\psi\phi$~\cite{conventional}. Thus, this structure does not 
fit conventional expectations for a charmonium state. 
We note that this structure 
decays to $J/\psi \phi$ just above
       the $J\psi \phi$ threshold, similar to the $Y(3930)$~\cite{y3940}, which decays to
       $J/\psi \omega$ near the $J/\psi \omega$ threshold. We therefore term it $Y(4140)$.

\par 
We thank the Fermilab staff and the technical staffs of the participating institutions for their vital 
contributions. We wish to thank E. Eichten for helpful discussions.
This work was supported by the U.S. Department of Energy and National Science Foundation; the 
Italian Istituto Nazionale di Fisica Nucleare; the Ministry of Education, Culture, Sports, Science and Technology 
of Japan; the Natural Sciences and Engineering Research Council of Canada; the National Science Council of the 
Republic of China; the Swiss National Science Foundation; the A.P. Sloan Foundation; the Bundesministerium f\"ur 
Bildung und Forschung, Germany; the Korean Science and Engineering Foundation and the Korean Research Foundation; 
the Science and Technology Facilities Council and the Royal Society, UK; the Institut National de Physique 
Nucleaire et Physique des Particules/CNRS; the Russian Foundation for Basic Research; the Ministerio de Ciencia e 
Innovaci\'{o}n, and Programa Consolider-Ingenio 2010, Spain; the Slovak R\&D Agency; and the Academy of Finland.

%\end{linenumbers}


\begin{thebibliography}{99}



\bibitem{x3872discovery}
S.-K.~Choi {\em et al.}~(Belle Collaboration),  \prl {\bf 91}, 262001 (2003);
D.~Acosta {\em et al.}~(CDF Collaboration), \prl {\bf 93}, 072001 (2004);
V.~M.~Abazov {\em et al.}~(D0 Collaboration), \prl {\bf 93}, 162002 (2004);
B.~Aubert {\em et al.}~(\babar~ Collaboration), Phys.\ Rev.\ D {\bf 71}, 071103(R) (2005).


\bibitem{y3940}
S.-K.~Choi {\em et al.}~(Belle Collaboration), \prl {\bf 94}, 182002 (2005);
B.~Aubert {\em et al.}~(\babar~ Collaboration), \prl {\bf 101}, 082001 (2008).



\bibitem{y4260discovery}
B.~Aubert  {\em et al.}~(\babar~ Collaboration), \prl {\bf 95}, 142001 (2005).


\bibitem{y4320}
B.~Aubert  {\em et al.}~(\babar~ Collaboration), \prl {\bf 98}, 212001 (2007);
X.~L.~Wang  {\em et al.}~(Belle Collaboration), \prl {\bf 99}, 142002 (2007);
S.-K.~Choi  {\em et al.}~(Belle Collaboration), \prl {\bf 100}, 142001 (2008);   
R.~Mizuk {\em et al.}~(Belle Collaboration),  Phys.\ Rev.\ D {\bf 78},  072004 (2008);
P.~Pakhlov {\em et al.}~(Belle Collaboration), \prl {\bf 100}, 202001  (2008).



\bibitem{conventional}
E.~Eichten, K.~Lane, and C.~Quigg, Phys.\ Rev.\ D {\bf 69},  094019 (2004);
E.~Eichten, S.~Godfrey, H.~Mahlke, and J.~Rosner, Rev.\ Mod.\ Phys. {\bf 80}, 1161 (2008).


\bibitem{hybrid}
S.~L.~Zhu, Phys.\ Lett.\ B {\bf 625}, 212 (2005);
F.~Close  and P.~Page, Phys.\ Lett.\ B {\bf 628}, 215 (2005).



\bibitem{molecule}
E.~S.~Swanson,  Phys.\ Lett.\ B {\bf 588},  189 (2004).


\bibitem{fourquark}
L.~Maiani, F.~Piccinini, A.~D.~Polosa, and V.~Riquer, Phys.\ Rev.\ D {\bf 72}, 031502(R) (2005);
D.~Ebert, R.~N. ~Faustov, and V.~O. ~Galkin, Eur. \ Phys.\ J.\ C {\bf 58}, 399 (2008).


\bibitem{x3872jpc}
A.~Abulencia {\em et al.}~ (CDF Collaboration), Phys.\ Rev.\ Lett.  {\bf 96}, 102002 (2006);
B.~Aubert {\em et al.}~(\babar~ Collaboration), Phys.\ Rev.\ D {\bf 74}, 071101 (2006).

\bibitem{ric}
N.~V.~Drenska, R.~Faccini, and A.~D.~Polosa, 	arXiv:hep-ph/0902.2803v1.


\bibitem{estiacharm07}
Yu.~S.~Kalashnikova and A.~V.~Nefediev, Phys.\ Rev.\ D {\bf 77}, 054025 (2008);
E.~J.~Eichten, Private communication;
P.~Hasenfratz, R.~R.~Horgan, J.~Kuti, and J.~-M.~Richard, 
Phys.\ Lett.\ B {\bf 95}, 299 (1980);
S.~Perantonis and C. Michael, Nucl.\ Phys.\ {\bf B347},  854 (1990);
C.~Bernard  {\em et al.}, Phys.\ Rev.\ D {\bf 56},  7039  (1997);
X. Liao and T. Manke, arXiv:hep-lat/0210030.



\bibitem{hybridfromb}
F.~E.~Close {\em et al.}, Phys.\ Rev.\ D {\bf 57},  5653 (1998); 
F.~E.~Close and S.~Godfrey, Phys.\ Lett.\ B {\bf 574}  210 (2003).

\bibitem{stanquarkonium}
S.~J.~Brodsky, I.~Schmidt, and G.~T\'eramond, \prl {\bf 64}, 1011 (1990).



\bibitem{cleojpsiphi}
A.~Anastassov  {\em et al.}~(CLEO Collaboration),  \prl {\bf 84}, 1393 (2000);
B.~Aubert {\em et al.}~(\babar~ Collaboration), \prl {\bf 91}, 071801 (2003). 


\bibitem{cdfii}
D.~Acosta  {\em et al.} (CDF Collaboration), Phys.\ Rev.\ D {\bf 71},  032001 (2005);
A.~Abulencia {\em et al.} (CDF Collaboration), Phys.\ Rev.\ Lett. {\bf 97}, 242003 (2006).


\bibitem{eta}
We use a coordinate system in which the proton beam
      direction is defined as the $z$ axis.  
The angle $\theta$ is the usual
   polar angle.  We define the
   pseudorapidity as  $\eta\equiv -\ln(\tan\frac{\theta}{2})$.  
   The transverse momentum is defined as  $p_T =p\sin\theta$, where $p$ is the momentum measured in 
   the tracking system.


\bibitem{PDG}
C.~Amsler {\it et al.}  ~(Particle Data Group), Phys.\ Lett.\ B {\bf 667}, 1 (2008).


\bibitem{cdfbslifetime}
T.~Aaltonen,  {\em et al.} (CDF Collaboration), Phys.\ Rev.\ Lett. {\bf 100}, 161802 (2008);
D.~Acosta  {\em et al.} (CDF Collaboration), Phys.\ Rev.\ Lett. {\bf 94}, 101803 (2005).


\bibitem{bsmix}
A.~Abulencia {\em et al.} (CDF Collaboration), Phys.\ Rev.\ Lett. {\bf 97}, 242003 (2006).


\bibitem{cdfbc}
A.~Abulencia {\em et al.} (CDF Collaboration), Phys.\ Rev.\ Lett. {\bf 96}, 082002 (2006);
T.~Aaltonen {\em et al.} (CDF Collaboration), Phys.\ Rev.\ Lett. {\bf 100} 182002 (2008).


\bibitem{pbw}
B.~Aubert  {\em et al.}~(\babar~ Collaboration), Phys.\ Rev.\ D {\bf 78},  071103 (2008).


\bibitem{sbw}
$\frac{dN}{dm} \propto \frac{m \Gamma(m)}{(m^2-m_0^2)^2+m_0^2\Gamma^2(m) }$, where
$\Gamma(m)=\Gamma_0\frac{q}{q_0}\frac{m_0}{m}$, and the 0 subscript indicates the 
value at the peak mass.


\bibitem{k1770}
T.~Armstrong {\em et al.}, Nucl.\ Phys.\ {\bf B221}, 1 (1983).


\end{thebibliography}
\end{document}%%%